# New and emerging forms of data and technologies: literature and bibliometric review

Petar Radanliev [1] 🔟 · David De Roure [1] 🔟

## Abstract

With the increased digitalisation of our society, new and emerging forms of data present new values and opportunities for improved data driven multimedia services, or even new solutions for managing future global pandemics (i.e., Disease X). This article conducts a literature review and bibliometric analysis of existing research records on new and emerging forms of multimedia data. The literature review engages with qualitative search of the most prominent journal and conference publications on this topic. The bibliometric analysis engages with statistical software (i.e. R) analysis of Web of Science data records. The results are somewhat unexpected. Despite the special relationship between the US and the UK, there is not much evidence of collaboration in research on this topic. Similarly, despite the negative media publicity on the current relationship between the US and China (and the US sanctions on China), the research on this topic seems to be growing strong. However, it would be interesting to repeat this exercise after a few years and compare the results. It is possible that the effect of the current US sanctions on China has not taken its full effect yet.

**Keywords** New and emerging forms of data · Literature review · Bibliometric review · Spatiotemporal data · Time-stamped data · Open data · Real-time data · High-dimensional data

✉ Petar Radanliev
   petar.radanliev@oerc.ox.ac.uk

[1] Oxford e-Research Centre, Department of Engineering Sciences, University of Oxford, Oxford, UK

# 1 Introduction

Traditionally, multimedia was related to a form of communication that combines data forms such as text, audio, images, animations of video in an interactive presentation. In this article, the term multimedia is related to new and emerging forms of interactive communications, build upon new and emerging types of data, analytics and visualisation techniques.

Traditional media data structures followed a simple and purpose specific design, storing relational databases in a tabular row and column. Modern multimedia data structures follow polymorphic design, where data entries from IoT apps, web sites, social networks, mobile devices, enhanced with artificial intelligence (AI), is coupled with object-oriented programming. Polymorphous data structures, stored as objects with nested elements, can change fast when new analytical features are built in dynamic new media for data analytics.

Big data usually includes terabytes or even zettabytes of structured, semi-structured and unstructured data, generated by human-machine interactions and processes. AI with deep learning (DL) and machine learning (ML) algorithms can analyse big data and derive understanding on events that occurred, and/or as they happen in real-time, and even forecast future events, enabling decisions on strategic directions. Traditional relational databases would struggle to fit, capture, manage, and process big data with low latency. There are many examples of when low-latency is desired [12], e.g. low-latency online gaming enables more realistic environment e.g., metaverse, or in high-frequency trading where trades are executed automatically with optimised algorithms capturing changing market prices. More recent example is the use of different media (e.g., wearables, social platforms) for the COVID-19 digital monitoring and decision making.

The aim of this article is to present a comparative study of qualitative and quantitative review of new and emerging forms of data and the application of such data with new and emerging technologies. The objective is to identify recent state-of-the-art in spatiotemporal data, time-stamped data, open data, real time data, and high-dimensional data. The second objective is to identify how new types of data are used with AI (ML/DL). The gap identified and researcher in this article is the application of new technologies to new types of data, to resolve contemporary problems, such as Covid-19.

This article consists of a literature review section, which also includes a review of research methodologies. Followed by a bibliometric study, comparison of results between the qualitative review and the quantitative study. Results and analysis section, and a conclusion section.

# 2 Literature review – Qualitative survey of literature

## 2.1 Processes that generate new forms of data: Data acquisition

New and emerging forms of data (NEFD) are very different types of multimedia data (e.g. transactions, registrations, internet, tracking, image), that are broadly grouped into the NEFD category. NEFD are collected from different types of connectivity. The coverage and bandwidth capabilities of different types of connectivity, mean that some NEFD (e.g. 5G), have longer range but require high energy consumption at handsets, while other NEFD (e.g. 6LoWPAN and WirelessHART) cover short communication range [16]. This means that these technologies would not be sustainable in Industrial internet of things (IIoT), because of low are coverage, or high energy consumption. But the internet of things (IoT) communications are based on a very diverse set of multimedia data sharing:

- Low Power Wide Area Networks (LPWANs) is a long-range communication that relies on small, cheap batteries that can last for years. LPWANs can connect to most IoT sensors, and are used for remote monitoring, smart metering, and many other functions that can operate with small blocks of data at a low rate. There are licensed LPWANs (NB-IoT, LTE-M) and unlicensed LPWANs (e.g. MYTHINGS, LoRa, Sigfox etc.).
- Mobile (5G) is a high-speed and ultra-low latency connection, best suited for autonomous vehicles and augmented reality, real-time video surveillance, real-time mobile connected health, and time-sensitive industrial automation.
- Zigbee wireless standard (IEEE 802.15.4) is used in short-range (<100 m), low-power, mesh protocol, (similar to e.g. Z-Wave, Thread etc.).
- Bluetooth Low-Energy (BLE) is commonly integrated into fitness and medical wearables and smart home devices and data is visualised on smartphones.
- Radio Frequency Identification (RFID) uses radio waves to transmit small amounts of data within a very short distance. RFID is mostly used in retail, logistics and supply chains.

In Table 1, the LPWAN (Sigfox, LoRa, Narrowband IoT (NB-IoT)) is most represented technology. LPWAN are low power technologies, with communications range from 1 km to 10 km, and NB-IoT node can last up to 10 years on a single battery life, and support 52,547 connections [16]. The weakness of LPWAN technology is the low data rate of up to 250kps, hence its most suitable as a complimenting technology. With the rise of connected devices using different technologies, distributing code data is becoming an issue. Code data dissemination methods are even considering the idea of vehicles being used in smart cities as communication systems for maximising coverage at low cost [77]. Another issue is that analytical methods and challenges for managing NEFD are also very different. It seems inevitable that artificial intelligence will need to be deployed to resolve many of these issues, especially in smart cities [2]. Deep learning algorithms can process big data at speed and high accuracy, but when the number of hidden layers increases from six, the deep learning process cannot be solved [42]. Some research methods are discussed briefly in the text below, just to understand how specific data types require very different approaches for analytics.

## 2.2 Types of social science research done using NEFD

### 2.2.1 Social science research using NEFD from IoT devices

The IoT generates, stores and processes real time big data that is offloaded and executed in centralised and decentralised data centres, e.g. cloud and edge servers. One of the main challenges of this process is optimising the execution time (i.e. low latency) and lowering energy consumption. To resolve these challenges, a computational offloading

**Table 1** Wireless connections for IoT endpoints

| Communication Technologies: | LPWAN | 5G | Zigbee | BLE | RFID |
|---|---|---|---|---|---|
| IoT endpoints: | | | | | |
| IIoT | * | ** | ** | | |
| Smart meter | * | | | | |
| Connected health | * | | | * | |
| Smart agriculture | * | | | | |
| Wearables | ** | | | * | |
| Smart building | * | | ** | ** | |
| Tracking | ** | * | | | * |

*widely used

**somewhat used

method is proposed for IoT cloud-edge computing [82]. Similar methods are designed for improving energy efficiency and increasing trust in IoT data, based on avoidance of unnecessary and untrustworthy data [79]. Similar technological solutions are enhancing the role of big data analytics from IoT and the Industrial IoT (IIoT), with enhancements for (1) industrial time series modelling; (2) intelligent shop floor monitoring; (3) industrial microgrids; (4) monitoring machine health; (5) intelligent predictive and preventive maintenance [68].

## 2.3 New methods involving NEFD as a tool for public engagement

### 2.3.1 Social media data as an alternative to traditional survey data

There is an increasing interest in data from social media e.g. Twitter, being used to supplement and/or substitute survey data. Studies as early as 2008 showed a strong correlation between the 'sentiment of tweets containing the word "jobs" and survey-based measures of consumer confidence', but a more recent study on tweets as an alternative to survey responses, showed lack of evidence and concluded that the 2008 data was a 'chance occurrence' [13]. One of the possible explanations presented in the study was that Twitter has become mainstream and the text has evolved into a modern language that differentiates from the original sentiment. But even after adjustments for the language differences, the correlation from 2008 could not be reproduced for recent years.

Oher studies show alternative viewpoint. Different methodological approach, using Facebook and Twitter data, has shown that data collection from 1000 participants, without any survey data, can provide effective forecast scores for millions of people [83]. Twitter 'sentiment' has also been used to replace customer satisfaction surveys, and the social media approach was found advantageous over survey data [29]. Social media showed stronger potential than survey data, because of dynamic feedback showing reactions to new and product releases. The social media approach was determined to offer a continuous, automated and lower cost feedback process, that added new insights that were not recorded in annual survey data [29].

Social media data was also found as more beneficial than survey data in visitor monitoring of a national park in Finland [30]. The main benefits of social media data vs survey data are stated as the continuous real-time data on visitor's behaviour and preferences. Even without asking any questions, the geotagged social media content provides sufficient data about who the visitors are, their activities, where in the park they went, among many other things. Geotagged social media data was also found useful in finding spatio-temporal activity patterns in areas where visitor monitoring was not taking place.

### 2.3.2 Big data as an alternative to traditional survey data

Big data is less costly and readily available, while survey data collection is slow and expensive, but the most promising outcome is to integrate both data sources [44]. The argument for integrating the low cost big data with the survey abilities to address specific questions with precise official statistics [20]. Nonetheless, the traditional value of survey data is diminishing with the declining survey participation, and in the future, it might be just one element of information data, which would comprise of different sources, including records and big data [55]. There is an argument that the survey research method, and the discipline of survey methodology, have adapted to technological innovations and will continue to evolve [15]. Because they represent strong and adaptable tools, supported by established theory and extensive evidence. Some studies applied manual grounded theory approach for improvement of computationally data driven intensive theory [8]. It is possible than survey methods would also evolve into the digital world by applying similar approach.

### 2.4 Background study for ML and DL in healthcare

In recent years (especially since the emergence of Covid-19), ML/DL approaches have been used extensively in healthcare settings. In this section we list some of the unexpected emerging innovations from the Covid-19 pandemic. Tables 2 and 3.

## 3 Types of NEFD – Bibliometric (statistical) review of data records

### 3.1 Spatiotemporal data

Spatiotemporal data includes location and time of individual events, enabling analysis of how events change in physical locations, e.g. changes in population over time; tracking objects in motion. Spatiotemporal data can be used for identifying caused in the past of anomalies that occur or become visible in a different time, e.g. time stamp of product anomaly, compared with production time of the same product. Spatiotemporal data can be analysed for building interactive visual analytics and show abstract view of data points in similar and different regions. Spatiotemporal data can be used to analyse data from localised areas, or to map spatial distribution in individual regions or to compare regions. GeoBrick [61] is one integrative technique that does all that. Other studies use spatiotemporal data to predict: the urban flow using machine learning [81]; air quality interpolation and visualisation in real-time [46]; cloud-terminal 'SuperMap' big data engine [78]; climate summer temperature zones [80]; and even the cholera hotspots in Zambia [58]. Spatiotemporal data models are used in various studies, including self-sustainable IoT networks that harvest energy from cellular network [7], and vehicle trajectory data in smart city [85]. Figures 1, 2 and 3.

**Table 2** Background study/review of recent ML/DL approaches used in healthcare settings

| Solution | IoT devices | IoT Gateway | Algorithm | Health condition | Year | Authors |
|---|---|---|---|---|---|---|
| Ambient Living | Microcontrollers: NodeMCU, Arouino, | Zighee, Zwave, Wi-Fi or LoRa gateway | Blockchain | Covid-19, heart disease, and diabetes. | 2021 | [59] |
| IoT cyber-attacks and anomaly detection | N/A – IoT system failure as a result of denial of service, data type probing, malicious control, malicious operation, scan, spying and wrong setup | N/A | Logistic regression, support vector machine, decision tree, random forest, and artificial neural network. | N/A – Cyber-attacks and anomaly detection | 2019 | [28] |
| **Solution** | **Tools** | **Dataset** | **Algorithm** | **Health condition** | **Year** | **Authors** |
| Diagnose and treat Covid-19 | Automatic extraction of features from X-ray images | Collection of 4575 X-ray images, including 1525 images of Covid-19 | Convolutional neural network (CNN) and long short-term memory (LSTM) | Covid-19, pneumonia | 2020 | [35] |
| Predictive analytics on Covid-19 recovery | Predictive data mining | Epidemiological dataset of COVID-19 patients of South Korea | Decision tree, support vector machine, naive Bayes, logistic regression, random forest, and K-nearest neighbor | Covid-19 recovery | 2020 | [57] |
| Breast Cancer prediction | 10-fold cross validation | UCI machine learning repository | Support vector machine and K-Nearest neighbors | Breast Cancer | 2017 | [34] |
| Covid-19 detection | CT and X-ray samples | Data collected from medical imaging samples | Deep Neural Networks | Covid-19 | 2021 | [39] |
| Diabetes prediction | Five-fold cross-validation | Pima Indian Diabetes (PID) data | Deep Neural Networks | Diabetes | 2019 | [71] |
| Facial mask detection | Image Pre-processing | CCTV cameras | Deep Neural Networks | Covid-19 | 2020 | [66] |
| Heart Disease Prediction | Computational intelligence | Statlog and Cleveland heart disease dataset | Logistic regression, support vector machine, decision tree, deep neural network, naïve bayes, random forest, and k-nearest neighbor | Coronary Artery Heart Disease | 2020 | [5] |
| Classification of liver disorder | 10-fold cross validation | BUPA liver dataset | Random forests and artificial neural networks | Liver disorder | 2018 | [27] |
| Automatic Covid-19 detection system | EMCNet | X-ray images | Convolutional neural network, random forest, support vector | Covid-19 | 2021 | [72] |

**Table 2** (continued)

| Solution | IoT devices | IoT Gateway | Algorithm | Health condition | Year | Authors |
|---|---|---|---|---|---|---|
| | | | machine, decision tree, and AdaBoost | | | |
| Breast cancer prediction | Comparative Study | Wisconsin Breast Cancer dataset | Support vector machine, K-nearest neighbors, random forests, artificial neural networks and logistic regression | Breast cancer | 2020 | [36] |
| Covid-19 management | Literature review | IEEE Xplore, PubMed, Google Scholar, Research Gate, and Scopus | Linear Regression, Multi-Layer Perceptron, Vector Auto-Regression, | Covid-19 | 2021 | [67] |
| Covid-19 management | Review | Real-time data | | Covid-19 | 2020 | [4] |
| Covid-19 diagnosis | Combined architecture for computational intelligence | X-ray images | VGG19, DenseNet121, InceptionV3, and Inception-ResNetV2 | Covid-19 | 2020 | [3] |
| Accidental falls | 10-fold cross-validation | Accelerometers, gyroscopes, RGB cameras, radars | Convolutional Neural Network, Long Short-Term Memory, and Auto-encoder based systems | Fall detection e.g., elderly | 2020 | [37] |
| Human-Computer Interaction | Review | State-of-the-art | Informative comparison | Emotion recognition | 2021 | [40] |
| Human-Computer Interaction | EEG Channel Correlation | Pearson's Correlation Coefficients (PCC) of alpha, beta and gamma sub-bands | Convolutional Neural Network | Emotion recognition | 2021 | [41] |
| Human-Computer Interaction | Webcam of RGB subtracted images | Haar cascade classifier | YCrCb skin segmentation | Hand movement | 2020 | [38] |
| Cancer identification | Social-behavioural factors | UCI machine learning repository | Decision tree, random forest, and xgboost | Cervical cancer growth | 2021 | [1] |
| Monitor the electrical behaviours of the devices in real-time | Support Vector Machine and Decision Tree | KDD Cup 1999, SEQUOIA 2000 | Density-Based Spatial Clustering of Applications with Noise, | Protect electronic devices | 2019 | [18] |
| Hepatocellular Carcinoma | SMOTE | XGBoost classifier | Machine learning | Patient's survival prediction | 2021 | [23] |

All of the NEFD that are used in the remaining part of this study can be collected from the Web of Science core Collection. This can be done for any country in the world, e.g., UK or USA and could also be done on a cumulative global level.

**Table 3** NEFD and new technologies

| NEFD | Data marketplace | Cloud-edge | SCADA | MES | ERP | CRM | Healthcare |
|---|---|---|---|---|---|---|---|
| Blockchain | [6, 9, 43, 50] | | | | | | |
| AI/ML | [56, 73] | [42] | | | | | [19] |
| Smart cities | [62] | [2] | | | | | |
| Edge computing | [56] | [79, 82] | | | SnappyData[a] | | |
| Blockchain joint cloud | [32] | | | | | | |
| IIoT | [75] | | [68] | [68] | [68] | [68] | |
| Digital single market. | [25] | | | | | | |

In short, NEFD is found in almost all new technologies, as listed in Table 3.

[a] https://snappydatainc.github.io/snappydata/

## 3.2 Time-stamped data

Time-stamped data is common in collecting user behaviour data and contains sequences of even time when data point was captured, or processed time when data point was collected. Time-stamped data enables understanding, predicting, and estimating individual actions over time, though journey analysis, individuals' steps taken and changes in time and responses. Time-series databases require data storage that can serve the execution of smart infrastructure queries, but cloud-based time-series storage can be expensive. With the increasing computing power and memory in connected devices, time-series data storage and analytics can be moved to the edge with distributed hash tables (DHT) [48]. Edge computing enables different types of predictive analytics, including big data driven predictive catching in mobile wireless networks (MWN) at the wireless edge [10]. Different types of time-stamped data analytics (e.g. IoT, edge, fog and cloud [22]) can also be integrated to analyse the spatio-temporal mobility data

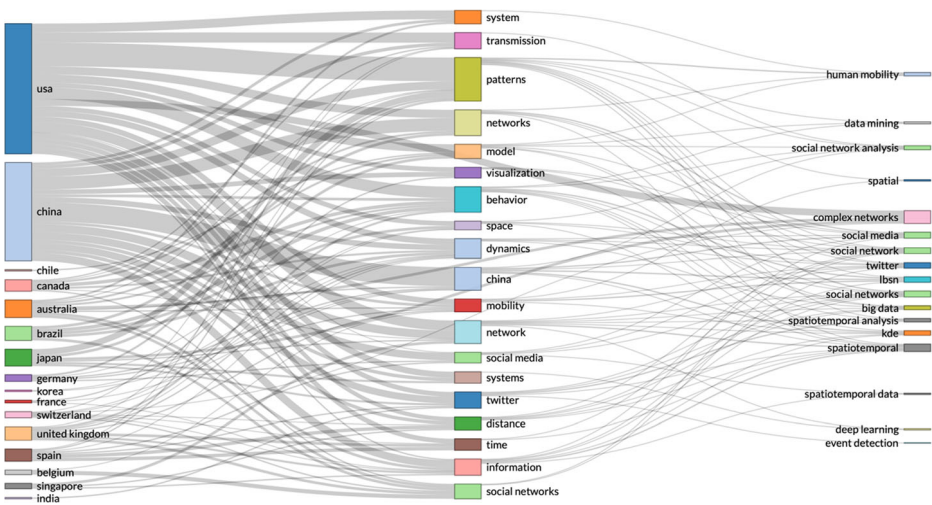

**Fig. 1** By country research output by key topics – search parameters (social networks AND spatiotemporal data)

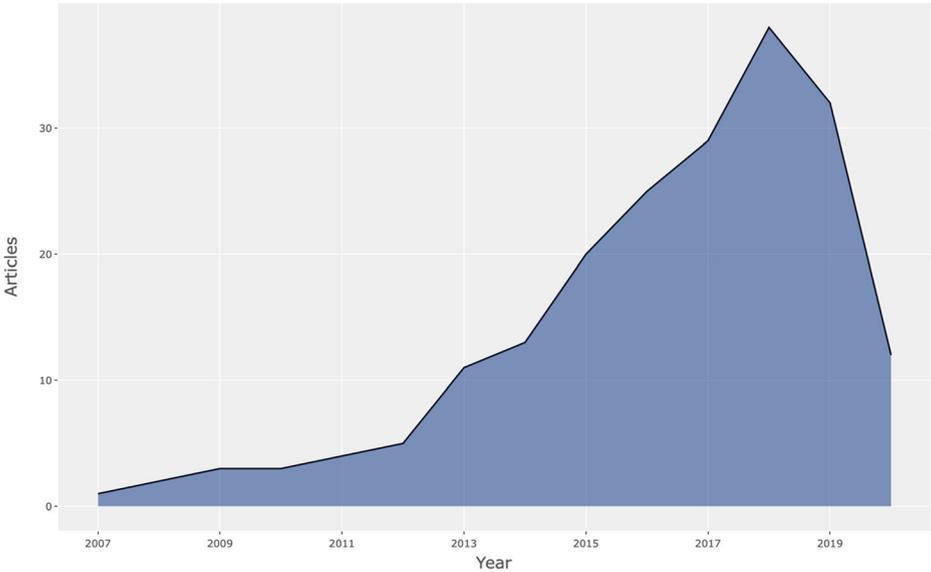

**Fig. 2** Same search parameters – number of publications are dropping in 2018, 2019 and 2020

(GPS logs), to predict locations of moving agents in real time, e.g. Mobi-IoST [24]. Edge computation is considered as speed-up over cloud computation, which could be of advantage in assessment based on street level data for on-edge traffic congestion [51]. Cloud computations on the other hand have proven abilities to obtain a *'time synchronization accuracy below 0.1 ms'* in estimating the accuracy of remote virtual machines for accurate time synchronisation in the Industrial IoT [69].

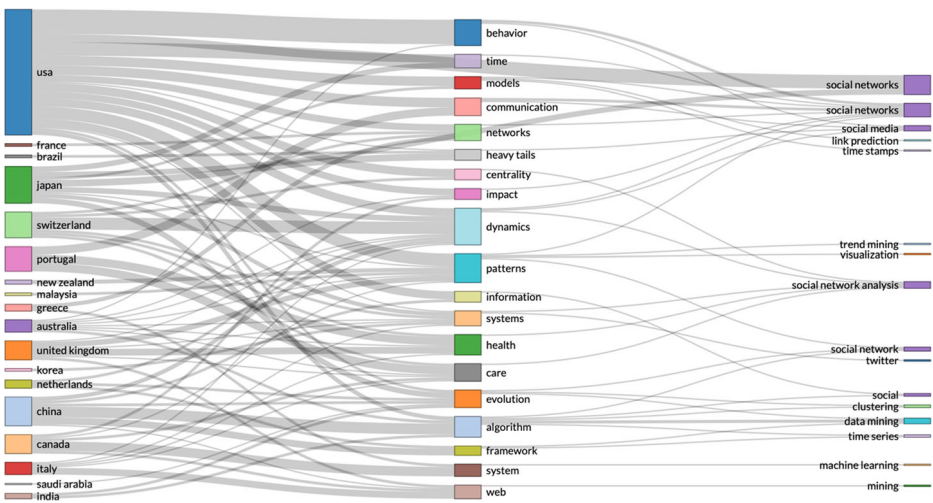

**Fig. 3** By country research output by key topics – search parameters (social networks AND time-stamped data)

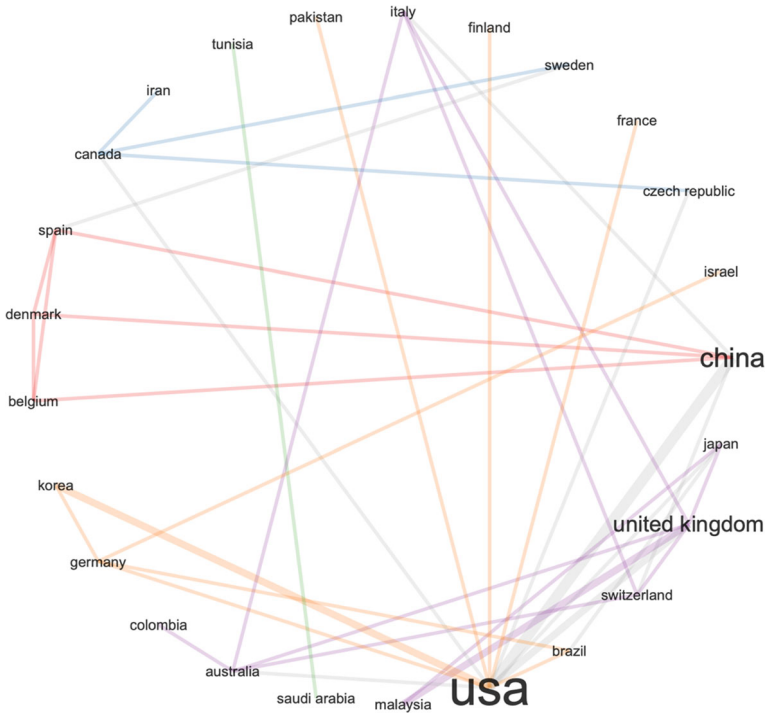

**Fig. 4** Same search parameters as in Fig. 3 – collaboration network

## 3.3 Open data

Open data is free for everyone to use, and it has been used for many different purposes. For example, Open Spending App[1] that tracks government spending worldwide in a standardised format, Elgin[2] provides real-time data on roadworks in the UK, and DataViva[3] delivers multiple visualisations on the Brazilian economy. There are many examples of open data usage. But so far, open aggregation of anonymised values, trends or analytics from private data, without disclosing sources, presented as open data, and used for commercial purposed, is only present in academic techniques. Given the increased EU wide regulations on personal data usage, tech companies that are built upon data security, seem reluctant to engage in public commercial usage of anonymised private data presented as open data. In addition, some data is extremely difficult to use, such as *unverified outdated data*. Unverified data is classified as collected data entries, without understanding of relevance, value, accuracy, or even if it's the correct data entry. One way of describing outdated data is when the evidence changes, but the data entries do not change. Using unverified outdated data could cause more damage than conducing analysis with no data Figures 4 and 5.

In addition to unverified outdated data, there is also dark data. *Dark data* is dormant data that is collected, processed, and stored during some form of regular activity, but it's not used

---

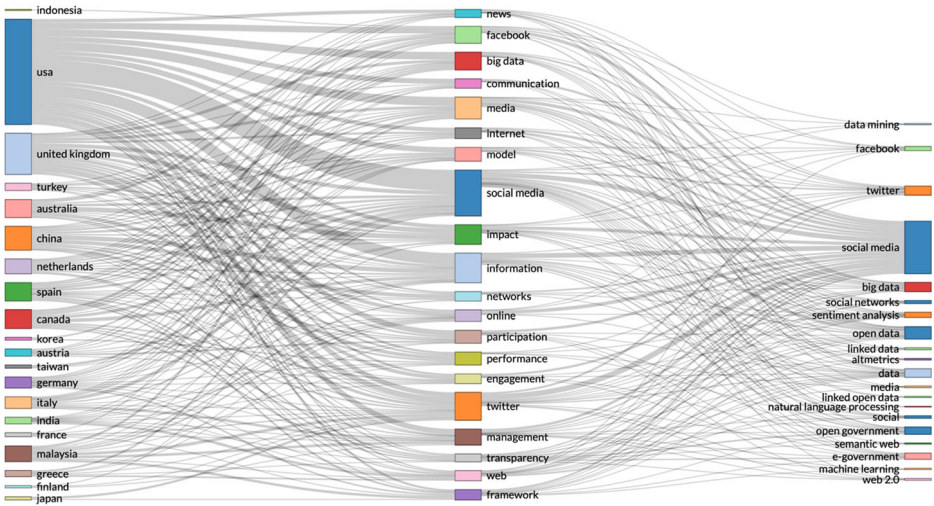

**Fig. 5** By country research output by key topics – search parameters (social networks AND open data) - from the 500 most relevant articles on WoS

for any purpose, and is kept as a dormant digital information asset. Dark data is often collected and stored as operational data. Companies collect and store application logs and metrics, events data, and information from third parties and microservices applications. Operational analytics is usually focused on how to turn that data into insights. The real question should be how to reverse engineer operational data metrics into a *data strategy* mindset of getting the right information from the start. Data strategy should include rethinking of what constitutes datasets and creating new possibilities for working with data. According to IBM, companies analyse only 1% of their data,[4] and over 60% of the data collected loses value immediately.[5] Dark data is different from unverified outdated data, but most of the dark data is unstructured and difficult to analyse.

The Open Data Institute[6] argues that open data is only useful if it's easy to understand, and can be tracked back to its origin, hence should be shared in standardised format. However, for open data to be more valuable, people need to trust in different interpretation of open data. Technologies such as the secure multi-party computation (also known as privacy-preserving computation),[7] and differential privacy,[8] could enable a broader interpretation, based on trust in privacy preserving technologies.

More work needs to be done on identifying and promoting the benefits for private companies to start sharing open data.[9] One example on how this could be taken forward is the PETRAS-IoT Data Management and Sharing Infrastructure (PEDASI) as a concept for a secure and legally trustworthy brokerage framework [26]. The PEDASI provides architecture for user, data and applications secure access to decentralised combined edge data sources. Similar privacy-preserving computation and differential privacy architectures can result with a significant increase in new data from IoT devices

---

and edge computing. Edge devices with high computational power can be enhanced with AI embedded autonomous and remote controlled edge processing, while low computational power, data could be sent to a server as an image, or through a real-time media stream [56].

Different example of promoting open data sharing is to identify open access machine data, and to promote the development of a 'digital single market' [25]. Machine data can be explained as a biproduct of everyday activities, e.g. data from mobile phone calls, driving connected cars, computer logs, etc., produced in unpredictable formats and often ignored. Although the term is in use for over 50 years, the rise of IoT brings new light for analysing data near real-time. Edge analytics for example enables automated analytical computation, at the point of collection.

### 3.4 Real-time data

Real-time data is becoming of ever-increasing relevance with the rise of edge analytics and 5G technology. The instant and immediate analytics of data creates value in different domains, mostly in the domain of smart cities. Real-time data is most valuable in crucial infrastructure e.g. emergency services, traffic control. But real-time data also has strong value in commercial events, e.g. marketing and advertising delivered at the precise moment, based on location and preferences Figure 6.

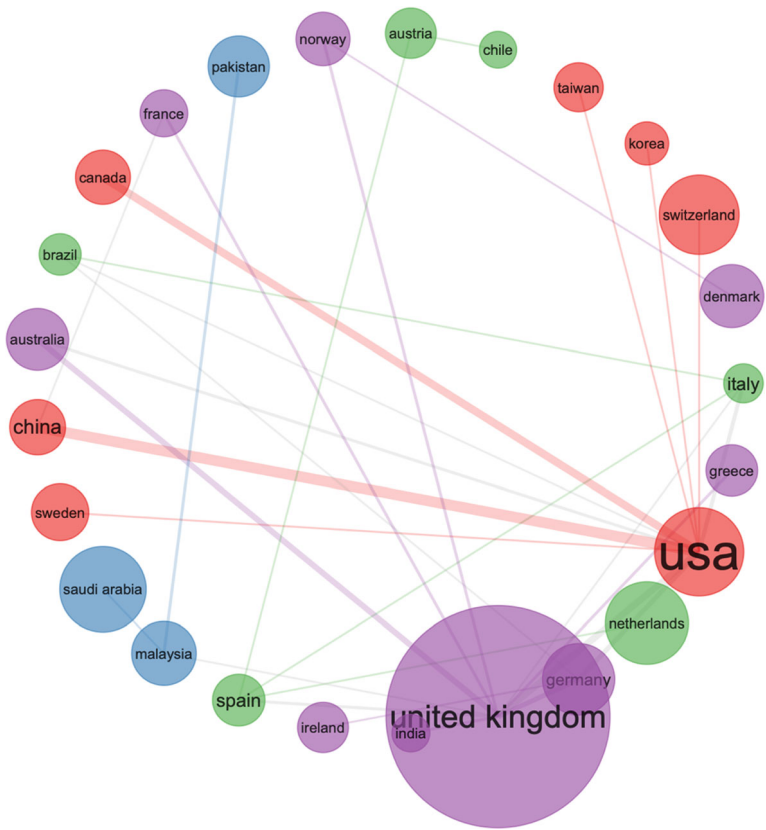

**Fig. 6** Same search parameters as in Fig. 5 – collaboration network – from the 500 most relevant articles on WoS

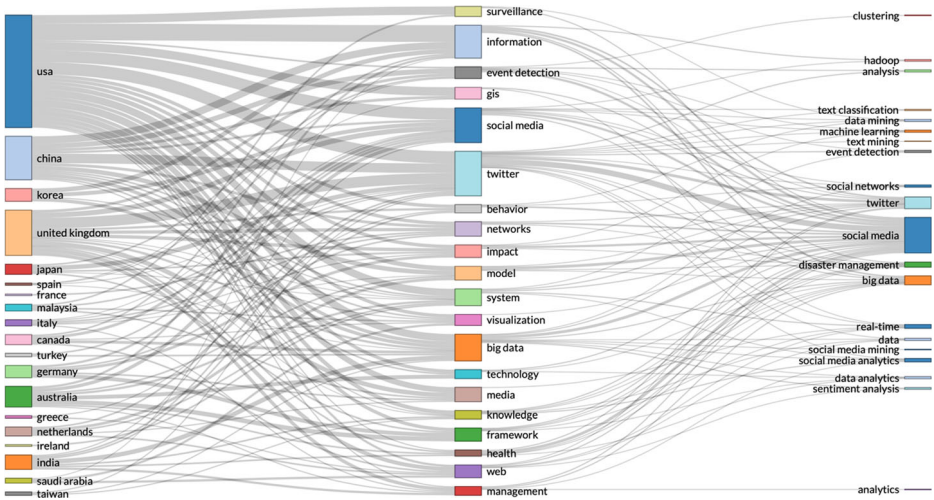

**Fig. 7** Analysis - UK seems more represented than China in the 500 most relevant articles on WoS

Real-time data is also used for monitoring and securing crucial systems producing high-dimensional data, where timely detection of abnormal data is crucial. For timely detection of abnormal data, a scalable algorithm is proposed that enables real-time nonparametric anomaly detection in high-dimensional settings [49], through *'Geometric Entropy Minimization'*. Figs. 7 and 8.

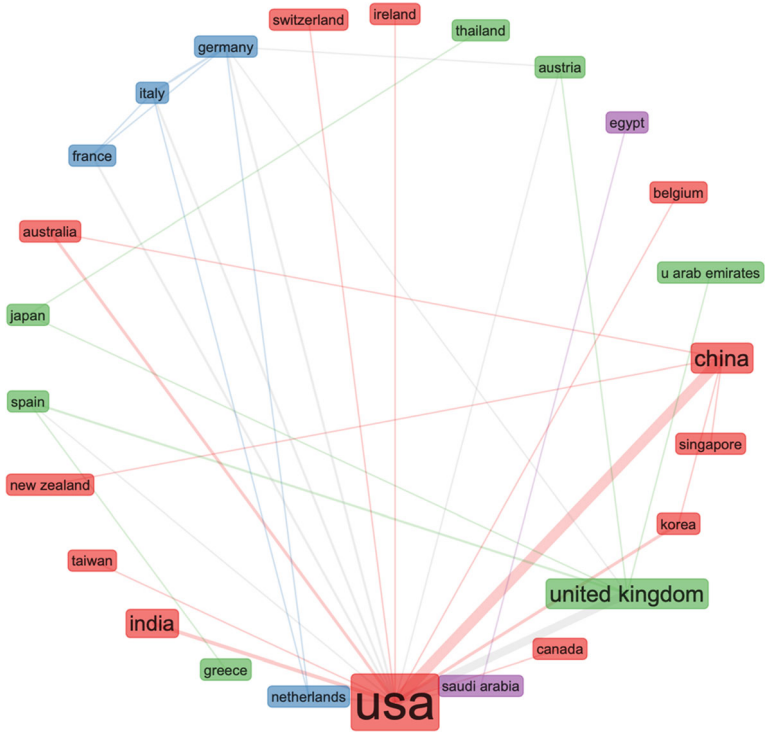

**Fig. 8** USA has strong presence in the collaboration network on real-time data analysis

### 3.5 High-dimensional data

High-dimensional data is defined as a high number of dimensions, in which the number of features exceeds the number of observations. For example, industrial high-dimensional big data reliability is studied and compared with a multi-method approach [52]. If we can analyse high-dimensional data, the research potential with is enormous. High-dimensional data has been used recently for IoT based smart farming in India [45], by applying improved genetic algorithm for extreme learning machine (IGA-ELM). High-dimensional data is also used in: finance, high resolution imaging, facial recognition technologies, etc.

High-dimensional data contains rich information, but also presents challenges in analysis and visualisation, while mapping high-dimensional data into lower-dimensional spaces often leads to information loss, unless multidimensional scaling (MDS) is performed [76]. A new data visualisation method called TMAP, enables visualisation of very large high-dimensional data sets as minimum spanning trees [63]. While for predictive analytics, a Bayesian framework for function-on-scalars regression proposed with many predictors [47].

The increased application of high-dimensional data from IoT, fog and cloud computing, has created many privacy concerns. Differential privacy preserving has emerged as a method to introduce noise and confuse adversaries by mixing sensitive input with noisy results. However, differential privacy has been criticised for poor utility and high complexity, caused by introducing noise to already complex high-dimensional data. A new compressed sensing mechanism (CSM) has been proposed to provide more accurate results [84], for differential privacy preserving. Figures 9 and 10.

## 4 Comparison results and discussion: An overview of the international landscape

The response to NEFD around the world is conflicted between the values and erosion of privacy. Some authors have expressed significant concerns about the appropriation of big data,

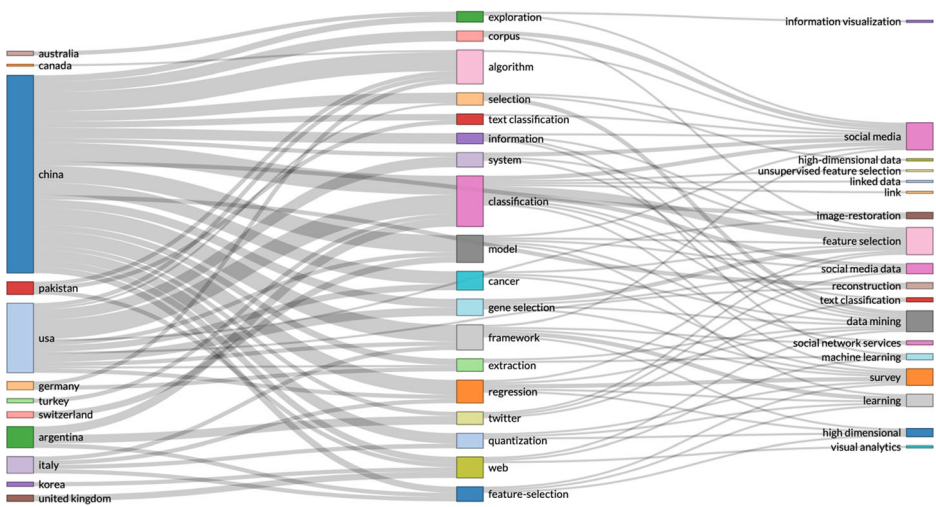

**Fig. 9** By country, China seems to lead in high-dimensional data analysis - from the most relevant 500 articles on WoS

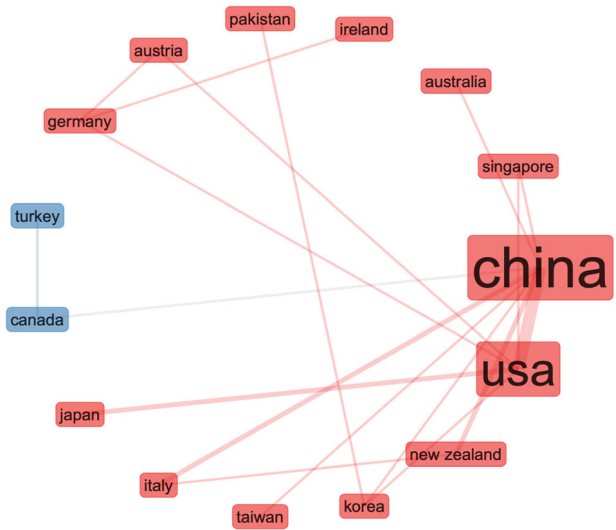

**Fig. 10** Collaboration network with at least one connection – analysis: China seems to be leading, followed by the US, while the UK is missing, 0 connections on this topic and the UK (from the 500 most relevant articles on WoS – search parameters: social media AND high-dimensional data)

and even compared the capture of social data to *'nothing less than a new social order, based on continuous tracking, and offering unprecedented new opportunities for social discrimination and behavioral influence'* [14]. Similar text call for disconnecting from the *'cybernetic loop'*, because even a google search reveals our intentions, and we should *'fight…against the rise of intelligent machines'* [31].

Other authors have focused on the more practical aspect and values of NEFD, such as smart health and monitoring, and promoted advancements towards 'autonomous wearable sensing for Internet of Things using big data analytics' [19]. Educational research has called for the implementation of big data in education [17].

The collaborative map in Fig. 11 shows very weak connection between the US and UK in this area. The results are derived by the following search terms (i.e. we searched the Web of Science Core Collection): ALL FIELDS: ("Emerging Data"). This search produced 147 results, refined by: RESEARCH AREAS: (AUTOMATION CONTROL SYSTEMS OR BEHAVIORAL SCIENCES OR COMPUTER SCIENCE OR ARTS HUMANITIES OTHER TOPICS OR INSTRUMENTS INSTRUMENTATION OR PHYSICS OR SOCIOLOGY OR MATHEMATICS OR INFORMATION SCIENCE LIBRARY SCIENCE OR TELECOMMUNICATIONS OR COMMUNI CATION). To check if these results are caused by the search terms, we searched the Web of Science Core Collection again for: ALL FIELDS: ("Emerging Data"), Timespan: Last 5 years, but without the filtering. This produced 1775 results. Figures 12 and 13.

### 4.1 IoT data marketplace

IoT data marketplaces with built in artificial intelligence, machine learning and edge computing enable device owners to sell their data [73]. IoT data is securely collected, stored, shared and sold in marketplaces,[10] with an increased focus on IoT data quality [53]. Concepts for blockchain enhanced global data marketplaces, based on smart contracts, are already in development [6, 9, 50]. IoT and blockchain are the two disruptive technologies associated most frequently in recent literature on smart cities [62]. Relevant questions have also been answered on the abilities of data providers to fulfil the data collection in on-demand decentralised marketplaces [60]. IoT data in marketplaces is usually distributed in real-time, or stored in the cloud for future sale, and cloud server storage creates a single point of failure. To prevent risk from this single point of failure, the concept of '*jointcloud*' is proposed, and data trade is supported with blockchain [32]. To achieve trust, transparency and non-reputability in a decentralised IoT marketplace with a limited trust, smart contracts can be used to mediate among trading brokers, data producers and consumers [6]. IoT data marketplaces are trying to securely monetarise data, and that requires development of strong reputation systems, hence the increased reference to blockchain in IoT data marketplaces [43]. The monetising of various IoT data, also requires different pricing mechanisms that ensure maximum value in different market settings [54]. Smart cities generate vast amount of sensor data among various different types of data from IoT devices captured in diverse data formats. This data needs to be transformed before sharing and loading [65]. Creating marketplace of services in smart community could be one method for synthesising and aggregating data resources that could be shared among different sets of open communities [21].

---

[10] https://data.iota.org/#/

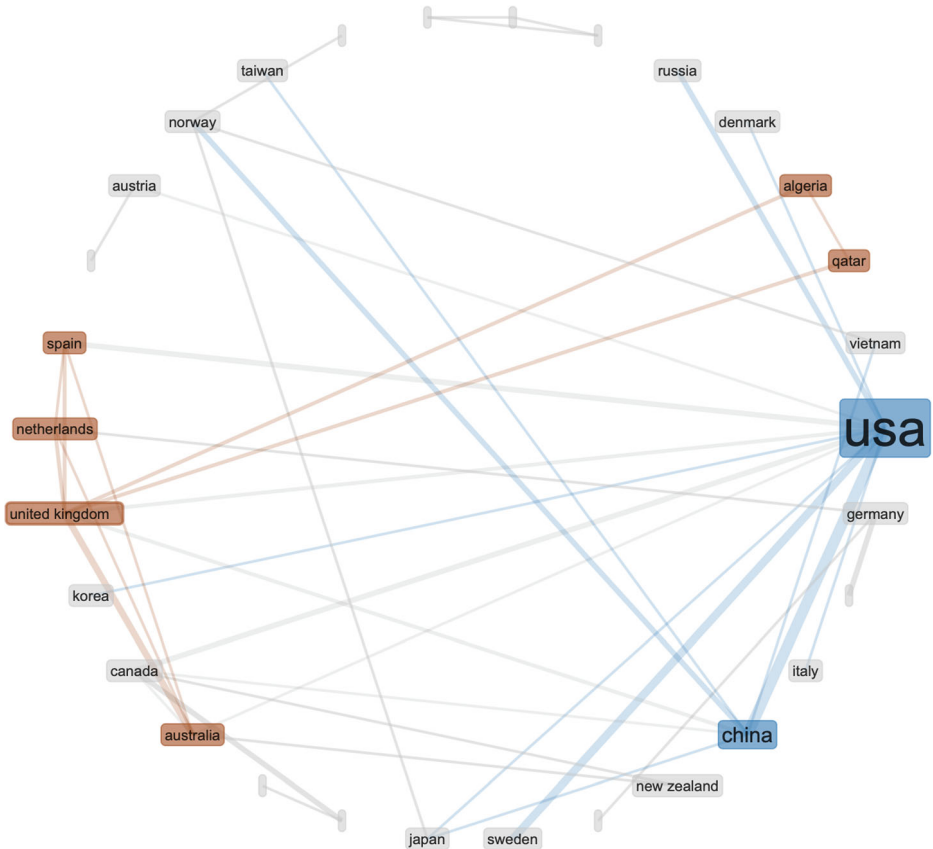

**Fig. 11** Surprisingly weak global collaboration with the US and UK (much stronger connecting with China)

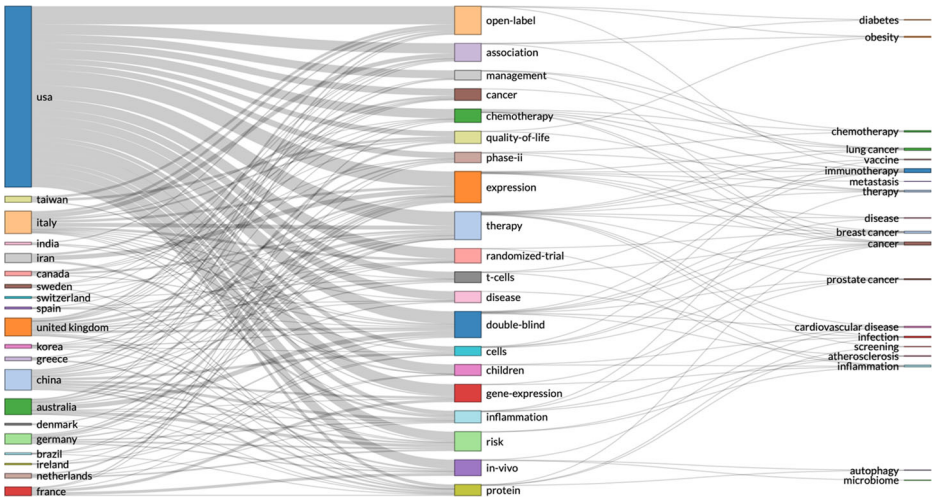

**Fig. 12** Analysis of the updated results - US seems to be leading in NEFD

However, there are ethical limits of blockchain enabled marketplaces for IoT private data, and a lack of careful consideration of the techno-economic impact, could result with the opposite effect, leading to *'the erosion of privacy for IoT users'* [33]. The concern is that even in a transparent private data market, we cannot be certain if the evaluation of diminishing data privacy reflects the established norms on privacy. The limitations of a traditional IoT and blockchain can be partially addressed with a permissioned demand driven analytics, enabling data democracy in the data supply chain [70].

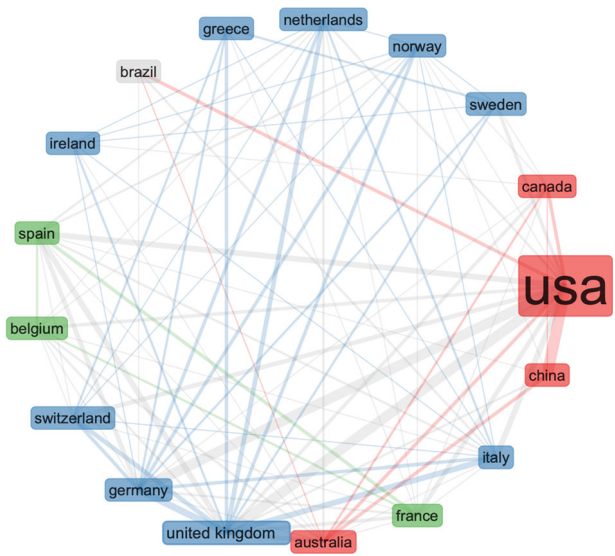

**Fig. 13** Different search, similar results like the previous Fig. 11 – weak collaboration between UK and US, – despite the news media coverage, the US seems to be working closer with China in this area

**Table 4** NEFD and security

| NEFD privacy preserving | IoT sensor data | IoT data marketplace | IIoT sensor data | References |
|---|---|---|---|---|
| Privacy preserving blockchain | x | | | [11, 70] |
| Blockchain data marketplace | | x | | [6, 9, 32, 43, 50, 60, 62] |
| Data property rights enforcement | | | x | [75] |

Blockchain technologies seem to be predominating the security literature. We found state-of-the-art solutions in privacy preserving, data preserving and rights enforcements.

## 4.2 Major NEFD research infrastructure initiatives and their impact

Existing enterprise based NEFD infrastructures include (1) supervisory control and data acquisition (SCADA); (2) manufacturing execution systems (MES); (3) enterprise resource planning (ERP) systems; (4) systems, customer relationship management (CRM) systems [68].

# 5 Results and analysis

## 5.1 Privacy-preserving data mechanisms for IoT data

Industrial IoT (IIoT) sensing as a service models suggest that industries are losing significant value from inadequate data sharing, which is usually caused by lack of property rights enforcement and relevant pricing models [75]. A new dynamic pricing mechanism, based on reinforcement learning, has been proposed for intelligent IoT data pricing [74]. Blockchain technology called 'sensor data protection system' (SDPS) has been proven successful in a tamper resistant IoT sensor data gathering, processing and exchange [11]. The main evaluation criteria for the SDPS is based on (1) tamper resistant in all stages of processing; (2) privacy preserving of the data owner; (3) capabilities to handle big data; and (4) economic feasibility of protection outweighing the cost. The use case study found that blockchain technology for security IoT sensor data, cannot assure tamper-proof without additional cross-validation, neither could assure data privacy without additional privacy such as IoT access control management [64]. Therefore, data protection should be implemented early in the processing, and hybrid blockchain is needed for certified data scaling, but economic feasibility is possible.

## 5.2 Findings on NEFD privacy preserving

With the rise of Bitcoin in 2009, we have seen a fast emergence of various blockchain technologies, which currently stand at over 17,000 on CoinMarketCap. [11]While some analysts have been sceptical of blockchains and crypto in the past, the volume of new technologies can no longer be ignored. Since Covid-19, the world has increased the adoption of new technologies, such as virtual reality and the metaverse, which support the further adoption of blockchain technologies. In Table 4 below, we list some of the recent security solutions based on blockchain technologies.

---

[11] https://coinmarketcap.com/

**Table 5** NEFD and applied solutions

| NEFD | Spatio - temporal data | Time -stamped data | Open data | Real-time data | High-dimensional data |
|---|---|---|---|---|---|
| | GeoBrick [61]. | Qubit.[a] | OSA[b]; Elgin[c]; DataViva.[d] | CUSUM [49]. | Industrial big data, multi-method [52]. |
| | Urban Flow prediction [81]. | Edge MWN [10]. | The Open Data Institute.[e] | Deep learning [42]. | IGA-ELM [45]. |
| | Air quality [46]. | Mobi-IoST [24]. | Multi-party computation[f]; Differential privacy.[g] | | MDS [76]. |
| | GIS platform [78]. | Edge DHT analytics [48]. | PEDASI [26]. | | TMAP [63]. |
| | ArcGIS [80]. | Cloud IIoT [69]. | Digital single market [25]. | | Bayesian regression predictors [47]. |
| | Cholera hotspots [58]. | | | | CSM [84]. |
| | RF energy [7]. | | | | |

With the increased number of data streams, one of the primary concerns are in the endpoint security and the increased cyber-attack surface from various data forms, connection protocols and edge devices.

[a] https://www.qubit.com/

[b] https://openspending.org/

[c] https://www.elgintech.com/

[d] http://dataviva.info/en/

[e] https://theodi.org/

[f] https://en.wikipedia.org/wiki/Secure_multi-party_computation

[g] https://en.wikipedia.org/wiki/Differential_privacy

## 5.3 Findings on NEFD analytics methods

The marketplace in different countries can be compared with the following NEFD categories – as listed in Table 5. We included some of the existing solutions, but we imagine these solutions are constantly on the rise. In not very distant future, we can imagine blockchain technologies contributing to the rise of digital decentralised marketplaces for NEFD.

## 6 Conclusion

This article conducted a literature review and bibliometric analysis on NEFD. The qualitative literature review engages with search of the most prominent journal and conference publications on this topic and produces some interesting insights on new methods involving NEFD as a tool for public engagement. The qualitative review outlines the processes that generate new forms of data (i.e. data acquisition) and reviews the types of social science research done using NEFD, including the social science research using NEFD from IoT endpoints. The qualitative review derives new methods involving NEFD as a tool for public engagement, with the use of social media data as an alternative to traditional survey data.

The second part of this article engages with statistical R analysis of Web of Science data records. The data records are searched for a few different types of NEFD data records, including spatiotemporal data, time-stamped data, open data, real-time data, and high-dimensional data. The results are somewhat surprising. By country and research output by key topics – (search parameters: social networks AND spatiotemporal data), confirms that the US and China are leading the research efforts in this area. But the number of publications are dropping in 2018, 2019 and 2020. Second unexpected results emerged from the analysis of country research output by key topics with a slightly edited search parameters: social networks AND time-stamped data. This analysis showed much lower research output from China, with US and Japan at the lead points. Since this result seems in conflict with the first result, we conducted further analysis of the same data records, and we discovered that US, China and UK have individual and isolated collaboration networks. Could this be interpreted that a lack of collaboration between the US and China affects the Chinese scientific output more than the US? We need further data to investigate that question.

The analysis of open data seems to provide some insights into this question. Considering the Chinese policies for open data sharing, we would expect China to be at the leading point of this area. However, the Web of Science data records show that the US is leading in scientific output on open data and the UK is leading in collaborative research on open data. Could this be caused by lack of regulatory compliance in analysing open data between different countries (e.g. EU- GDPR)? Again, we need further data to investigate that question. The statistical analysis of real-time data records presented closer collaboration between US and China, but it also showed that UK is performing strongly in this research area – while not collaborating strongly with either the US or China. The US research dominance changes in the area of high-dimensional data, where China seems to lead strongly – in research output and collaboration.

Finally, we conducted a statistical analysis of the international landscape of research collaborations on NEFD, and the result were again unexpected. The research collaboration between the US and China seems to be growing stronger over the past few years. To eliminate doubts and data bias, we conduced alternative search and we reanalysed the data – just to reach the same results. It appears that politics is not affecting science as much as we expected. However, scientific results (i.e. output) can often take many years and this results need to be reanalysed with updated data records in a few years' time, to check if the results remain the same.

**Acknowledgements**  Eternal gratitude to the Fulbright Visiting Scholar Project.

**Code availability**  N/A – no code was developed; code was however used for running the R Studio analysis.

**Authors contributions**  Both authors contributed equally. Both authors contributed to the study conception and design. Material preparation, data collection and analysis were performed by Dr. Petar Radanliev, and Professor David De Roure. The first draft of the manuscript was written by Dr. Petar Radanliev and both authors commented on previous versions of the manuscript. Both authors read and approved the final manuscript.

**Funding**  This work was funded by the ESRC [grant number:].

**Data availability**  all data and materials included in the article.

## Declarations

**Competing interests** On behalf of all authors, the corresponding author states that there is no conflict nor competing interest.

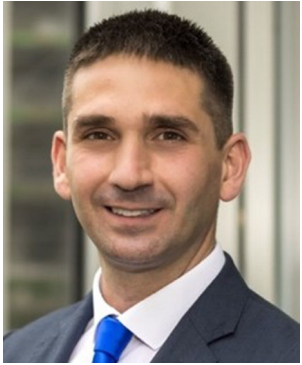

**Petar Radanliev** is a Post-Doctoral Research Associate at the University of Oxford. He obtained his Ph.D. at University of Wales in 2014 and continued with Postdoctoral research at Imperial College London, Massachusetts Institute of Technology, University of Cambridge and University of Oxford. His current research focusses on artificial intelligence, internet of things, and cyber risk analytics at the edge.

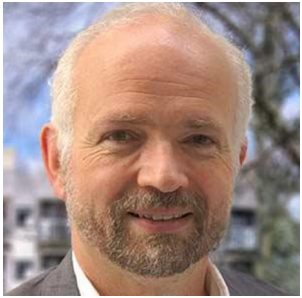

**David De Roure** is a Professor of e-Research at University of Oxford. He obtained his Ph.D. at University of Southampton in 1990 and went on to hold the post of Professor of Computer Science, later directing the UK Digital Social Research programme. His current research focusses on social machines, Internet of Things and cybersecurity. He is a Fellow of the British Computer Society and the Institute of Mathematics and its Applications.